%
%

\input harvmac.tex

\def\IR{\relax{\rm I\kern-.18em R}}
\def\IZ{\relax\ifmmode\mathchoice
{\hbox{\cmss Z\kern-.4em Z}}{\hbox{\cmss Z\kern-.4em Z}}
{\lower.9pt\hbox{\cmsss Z\kern-.4em Z}}
{\lower1.2pt\hbox{\cmsss Z\kern-.4em Z}}\else{\cmss Z\kern-.4em
Z}\fi}
\font\cmss=cmss10 \font\cmsss=cmss10 at 7pt

%

%



\lref\juan{J. Maldacena, ``The Large N Limit of Superconformal Field 
Theories and Supergravity'', hep-th/9711200}

\lref\witten{E. Witten, ``Anti de Sitter Space and Holography'', 
hep-th/9802150}

\lref\shamiteva{S. Kachru and E. Silverstein, ``4D Conformal Field 
Theories and Strings on Orbifolds'', hep-th/9802183}

\lref\horwitt{P. Horava and E. Witten, ``Heterotic and Type I Dynamics 
from Eleven Dimensions'', hep-th/9510209, Nucl. Phys. B460,506,1996}

\lref\oriami{O.J. Ganor and A. Hanany, ``Small E(8) Instantons and 
Tensionless Noncritical Strings'', 
hep-th/9602120, Nucl. Phys. B474,122,1996}

\lref\nawitta{N. Seiberg and E. Witten, ``Comments on String 
Dynamics in Six Dimensions'', hep-th/9603003, Nucl. Phys. B471,121,1996}

\lref\hirgar{G. Horowitz and H. Ooguri, ``Spectrum of Large N 
Gauge Theories from Supergravity'', hep-th/9802116}

\lref\igoste{S.S. Gubser, I. Klebanov and A. Polyakov, ``Gauge 
Theories Correlators from Non-critical String Theory'', hep-th/9802190}

\lref\igogub{S.S. Gubser, I.R. Klebanov and A.W. Peet, "Entropy and 
Temperature of Black 3-Branes", hep-th/9602135,
Phys. Rev. D54,3915,1996; S.S. Gubser and I.R. Klebanov, ``Absorption
by Branes and Schwinger Terms in the World Volume Theory'',
hep-th/9708005, Phys. Lett. B413,41,1997}

\lref\juantau{N. Itzhaki, J.M. Maldacena, J. Sonnenschein, 
S. Yankielowicz, ``Supergravity and the Large N Limit of Theories with 
Sixteen Supercharges'', hep-th/9802042}

\lref\kallosh{P. Claus, R. Kallosh, J. Kumar, P. Townsend and 
A. Van Proeyen, ``Conformal Theory of M2, D3, M5 and D1 Branes+D5 
branes'', hep-th/980150;
R. Kallosh, J. Kumar, A. Rajaraman, ``Special Conformal Symmetry of
World Volume Actions'', hep-th/9712073; P. Claus, R. Kallosh and
A. Van Proeyen, ``M 5-branes and superconformal (0,2) tensor Multiplet
in 6 Dimensions'', hep-th/9711161}

\lref\skenderis{H.J. Boonstra, B. Peeters, K. Skenderis, ``Branes and 
Anti-De Sitter Space-Time'', hep-th/9801076;
K. Sfetsos and K. Skenderis, ``Microscopic Derivation of the 
Bekenstein Hawking Entropy Formula for Nonextremal Black Holes'', 
hep-th/9711138;
H. J. Boonstra, B. Peeters, K. Skenderis, ``Duality and asymptotic
geometries'', hep-th/9706192, Phys.Lett. B411 (1997) 59}

\lref\minic{M. Gunaydin and D. Minic, ``Singletons, Doubletons and 
M-theory'', hep-th/9802047}

\lref\vafaf{C. Vafa, ``Evidence for F Theory'', hep-th/9602022, 
Nucl. Phys. B469,403,1996}

\Title{\vbox{\baselineskip12pt\hbox{hep-th/9802195}
\hbox{IASSNS-HEP-98/18}}}
{\vbox{\centerline{A Supergravity Dual of a (1,0) Field Theory}
\centerline{}
\centerline{in Six Dimensions}}}

\centerline{Micha Berkooz$^1$} 
\smallskip
\smallskip
\smallskip
\smallskip
\centerline{$^1$ Institute for Advanced Study}
\centerline{Princeton, NJ 08540, USA}
\centerline{\tt berkooz@ias.edu}
\bigskip
\bigskip
\noindent
We suggest a supergravity dual for the $(1,0)$ superconformal field
theory in six dimensions which has $E_8$ global symmetry. Compared to
the description of the (2,0) field theory, the 4-sphere is replaced by
a 4-hemisphere, or by orbifolding the 4-sphere.

\Date{February 1998}


\newsec{Introduction}

It was recently conjectured by \juan\ that certain string theory or
M-theory backgrounds are dual to the large N limit of field
theories. The supergravity solution has therefore to reflect all the
symmetries of the field theory. A particular example is that of
superconformal field theories. For a theory in $d+1$ dimensions the
conformal symmetry is $SO(d+1,2)$. On the supergravity side one sees
this symmetry geometrically, as the background is\foot{At least in
some cases.} a product $AdS_{d+1}\times K$, where $K$ is a compact
manifold, on which other global symmetries of the theory may act.

A particular example of such a solution is that of the $(2,0)$ field
theory in 5+1 dimensions. The supergravity description is given by
M-theory compactified on $AdS_7\times S^4$ \juan. The $AdS_7$ part
realizes the conformal symmetry and the $S^4$ realizes the $SO(5)$
R-symmetry of the theory. Further analysis of this correspondence, in
a related model, is carried out is
\refs{\witten,\hirgar,\igoste}. Other cases and related work appears
in \refs{\juantau, \kallosh, \skenderis, \minic, \igogub}.

Another way of obtaining global symmetry is described in \witten. If
there are gauge fields on the supergravity side then these naturally
couple to symmetry currents on the boundary, and hence the field
theory will have additional global symmetries. It was shown in
\witten, that one obtains the correct 2-pt functions for the global
symmetry currents.

In the following short note we will discuss a realization of global
symmetry in such a way, in the context of a $(1,0)$ field theory with
$E_8$ symmetry in 5+1 dimensions.

Related work also appears in \shamiteva.

\newsec{The $(1,0)$ field theory in 5+1 dimensions}

One way of realizing this $(1,0)$ field theory is in M-theory on
$S^1/Z_2$ \horwitt. The field theory is then the low-energy limit of
the excitations on an M-theory 5-brane that is located at the fixed
point of the $Z_2$ action \refs{\oriami,\nawitta}, and is in a
superconformal fixed point. This information is sufficient to
determine the M-theory background that corresponds to this field
theory when the number of 5-branes, N, is taken to infinity.

In the case of $AdS_7\times S^4$, one set of coordinates on the
$AdS_7$ is $X^{0..5}$ and $U$. These are the coordinates that one
naturally gets when starting from the 5-brane solution which is
asymptotically flat and then performing the scaling procedure in
\juan. The $X$'s parametrize coordinates parallel to the M5-brane and
$U$ is a radial coordinate away from it. $S^4$ then parametrizes the
angles around the M5-brane.

The same seems roughly also to apply to the case of an M5-brane at the
end of the world. Let us denote the coordinates that the 5-brane spans
by $X^{0..5}$ and the coordinate of $S^1/Z_2$ by $X^{10}$. Since the
theory is still conformal one has to obtain an $AdS_7$ as part of the
solution. When we try and obtain the solution from the the solution
which is asymptotically flat, the coordinate $U$ then corresponds to a
combination of the distance in the $X^{10}$ direction and in the
$X^{6..9}$ directions. The internal manifold $K$, which replaces the
$S^4$ in the case at hand, parametrizes the directions on the
hypersurface $U=Const$. The 9-brane correspond to a boundary of $K$.

We are further restricted by the global symmetries. The theory has an
$SO(4)$ symmetry. This implies that the 4-dimensional manifold is made
of a bundle of 3-sphere over a 1-dimensional manifold. The boundary is
also a 3-sphere. Since we expect this manifold to be compact (we do
not want additional low-lying excitations other than the ones on the
$AdS_7$) the only candidate in a hemisphere.

Away from the boundary, we can also calculate the metric. The boundary
started its life as the boundary in M theory on $S^1/Z_2$. As it is
not a source for any field strength, it does not affect the solution
away in the bulk. This implies that the metric is also that of a
hemisphere.

In short, to obtain the supergravity descirption of the $(1,0)$ field
theory, one orbifolds the $S^4$. This also teaches us how to treat the
fixed points, i.e., the boundary of the hemisphere. As we need to keep
all the degrees of freedom that appear in low-energy
supergravity\foot{In this case we can not hope to do more then a low
energy analysis.}, one need to place, as in \horwitt, an $E_8$ worth
of vector fields on the boundary. The $E_8$ gauge fields therefore
propagate throughout the $AdS_7$. The field theory then has a global
$E_8$ symmetry, and the gauge fields in supergravity correctly
reproduce at least some aspects of current-current correlators.
   
To completely specify the background one needs to specify the state of
the gauge field degrees of freedom on $AdS_7\times S^3$. As we are
interested in the point where the $E_8$ symmetry is unbroken, the
configuration is such that none of the $E_8$ charged fields have an
expectation value. This also solves the equations of motion that
couple the bulk to the boundary.

\newsec{Discussion}

One expects that this procedure can be generalized to other cases, in
precisely the same way by which brane probes acquire a global symmetry
G when approaching a locus in spacetime where there are localized
gauge degrees of freedom of G (such as in F-theory \vafaf). In the
case of conformal field theories one expects that there will be gauge
bosons which will be smeared on the entire $AdS$ and localized at
hypersurfaces in $K$.

In particular, the dual supergravity/string theory description is
useful in a regime where the curvature is small and when string
perturbation theory is reliable. In that case, then when we are close
to the singularity in the supergravity side, the correct way to treat
the singularity is in string perturbation theory (or by a set of
non-perturbative techniques). Because the curvature is small, one
expects that its qualitative features will not change from those of
such a singularity in flat space\foot{At least locally. Globally we
could have additional effects. These may corresponds to effects due to
the flow, such as spontaneous symmetry breaking, or to a change of
parameters in the field theory.}. In particular, if we obtain enhanced
gauge symmetry from a singularity in flat space, one expects it to
persist in the regime where perturbative string theory around a nearly
flat space is a useful approximation. These gauge bosons will reflect
the global symmetry of the field theory, as we have seen for the case
of the $(1,0)$ field theory above.

\vskip 1cm

\centerline{\bf Acknowledgments}\nobreak

We would like to thank N. Seiberg, M. Strassler and E. Witten for
useful discussions. This work is supported by NSF grant NSF
PHY-9512835.

\listrefs

\end